\documentclass{aa}  

\usepackage{graphicx}
\usepackage{txfonts}
\usepackage{natbib,twoopt}
\usepackage[breaklinks=true]{hyperref} 
\bibpunct{(}{)}{;}{a}{}{,}             
\makeatletter
  \newcommandtwoopt{\citeads}[3][][]{\href{http://adsabs.harvard.edu/abs/#3}%
    {\def\hyper@linkstart##1##2{}%
     \let\hyper@linkend\@empty\citealp[#1][#2]{#3}}}
  \newcommandtwoopt{\citepads}[3][][]{\href{http://adsabs.harvard.edu/abs/#3}%
    {\def\hyper@linkstart##1##2{}%
     \let\hyper@linkend\@empty\citep[#1][#2]{#3}}}
  \newcommandtwoopt{\citetads}[3][][]{\href{http://adsabs.harvard.edu/abs/#3}%
    {\def\hyper@linkstart##1##2{}%
     \let\hyper@linkend\@empty\citet[#1][#2]{#3}}}
  \newcommandtwoopt{\citeyearads}[3][][]%
    {\href{http://adsabs.harvard.edu/abs/#3}
    {\def\hyper@linkstart##1##2{}%
     \let\hyper@linkend\@empty\citeyear[#1][#2]{#3}}}
\makeatother

\begin{document}

   \title{Searching for orbital period modulation in X-ray observations of the symbiotic X-ray binary GX 1+4}

   \author{Moritz Klawin\inst{1}\and
          Lorenzo Ducci\inst{1,2,3} \and M. Mirac Serim\inst{1}\and Andrea Santangelo\inst{1}\and Carlo Ferrigno\inst{2,3} \and Enrico Bozzo\inst{2}}

   \institute{$^1$Institut f\"ur Astronomie und Astrophysik, 
              Sand 1, 72076 T\"ubingen, Germany\\
              $^2$ISDC Data Center for Astrophysics, Universit\'e de Gen\`eve, 16 chemin d'\'Ecogia, 1290 Versoix, Switzerland\\
              $^3$INAF -- Osservatorio Astronomico di Brera, via Bianchi 46, 23807 Merate (LC), Italy\\
              \email{klawin@astro.uni-tuebingen.de}}

   \date{Received 28 August 2024; Accepted 02 September 2024 }

 
  \abstract{The symbiotic X-ray binary GX 1+4 possesses a number of peculiar properties that have been studied since the early 1970s. In particular, the orbital period has been a point of debate for many years, until radial velocity measurements were able to settle the debate. These radial velocity findings have so far not been confirmed using X-ray data, even though multiple factors would cause a periodic variation on the same timescale as the orbital period at these energies. Because the orbit of GX 1+4 is eccentric and not seen face-on, changes in the accretion rate and column density along the line of sight could cause a periodic variation in the spin-frequency measurements, X-ray light curves, and hardness ratios of the source. Furthermore, for a high inclination of the orbital plane, the neutron star could be eclipsed by the companion, which would lead to periodic decreases in brightness. We used data from a number of different X-ray telescopes to search directly for periodicity by applying the Lomb-Scargle and epoch-folding approaches to long-term light-curve and spin-frequency measurement data of the source. We support our findings using folded light curves, hardness ratios, and images. We find that our results agree with the radial velocity findings, and we form a self-consistent model that is supported by folded hardness-ratios and light curves. We find that the source is clearly detected in X-rays during the predicted eclipse. Motivated by this absence of an eclipse in the system, we constrain the inclination of the system to $\sim 76^\circ-84^\circ$ and the mass of the neutron star in the system to $\sim1.23M_\odot - 1.45M_\odot$ using the constraints on the red giant mass and surface gravity provided in the literature. Furthermore, we constrain the radius of the red giant to $\sim 60R_\odot - 150R_\odot$.}


   \maketitle
%

\section{Introduction}
Symbiotic binaries\footnote{For a comprehensive compilation of symbiotic variables see \url{https://sirrah.troja.mff.cuni.cz/~merc/nodsv/}} are systems consisting of a red giant star and a compact object \citep{1997A&A...319..201M}, which typically is a white dwarf \citep{1976Afz....12..521T}. In a handful of cases, however, the compact object has been confirmed to be a neutron star. These systems can be observed to emit in X-rays and are called symbiotic X-ray binaries (SyXRBs) \citep{2006A&A...453..295M}. At the time of writing, only five such systems have been confirmed (see, e.g., \citealt{2022MNRAS.513...42B} and references therein).
While these systems are a subclass of low-mass X-ray binaries (LMXBs), they exhibit characteristics that are often found in high-mass X-ray binaries (HMXB) instead, such as long pulse periods (up to hours), strong X-ray variability (factors of 10-20), and extremely long orbital periods of up to thousands of days \citep{2022MNRAS.513...42B}. Because these systems are rare and show a multitude of peculiar characteristics, they are prime candidates for further studies.

One of these rare objects is the SyXRB GX 1+4, which was first discovered in X-rays during balloon-borne experiments in 1970 \citep{1971ApJ...169L..17L} in the form of a new variable source with a pulsation period of $\sim$135s. Three years later, the optical counterpart of the source was identified by \citet{1973NPhS..245...39G} as V2116 Oph, a bright infrared source. This allowed \citet{1977ApJ...211..866D} to classify GX 1+4 as an SyXRB containing an M-giant mass donor. This made GX 1+4 the first of this class of SyXRBs. 
It is understood today that the system consists of a neutron star and a giant star companion \citep{1977ApJ...211..866D} that does not fill its Roche lobe \citep{2006ApJ...641..479H}, with evidence for an accretion disk in the system that is either persistent (\citealt{1997ApJ...482L.171J, 1997ApJ...489..254C, 1998ApJ...497L..39C}) or transient (\citealt{2017MNRAS.469.2509S}). The spectral type of the donor is either M6 III \citep{1977ApJ...211..866D} or M5 III \citep{1997ApJ...489..254C}. The constituent masses of the binary system are not conclusively determined, but \citet{2006ApJ...641..479H} suggested a maximum mass of the optical companion of $1.22M_\odot$ based on radial velocity measurements and adopting a typical neutron star mass of $1.35\mathrm{M_\odot}$. The neutron star showed a strong spin-up trend since its discovery. The spin period decreased from the initial value to $\sim$110s until 1980 \citep{1982MNRAS.201..759R}. The system then entered a faint state and remained undetectable \citep{1983IAUC.3872....1H} until 1986, when it was again detected after undergoing a torque reversal during the faint state and showed a spin-down trend with $\Dot{P} \sim 4.5 \cdot 10^{-8} \mathrm{s\,s^{-1}}$, which has stayed roughly constant since then (see \citealt{2012A&A...537A..66G} and references therein for a comprehensive overview of the spin history evolution). Newer measurements by \cite{2023A&A...676L...2L} suggested a spin-down of $\Dot{P} \sim 1.61 \cdot 10^{-7} \mathrm{s\,s^{-1}}$ based on data from the Kepler/K2 space telescope from observations carried out in 2016.

Multiple attempts have been made over the years to determine the orbital period of the system. Initially, \cite{1986ApJ...300..551C} found a periodicity of $\sim304$d based on epochs of enhanced spin-up in the spin-derivative history. Adopting an analogous approach, \citet{1999ApJ...526L.105P} inferred similar periodicities based on the residuals of the spin-frequency derivative history of the neutron star and connected it to the orbital period based on an increase in torque exerted by the accretion disk on the neutron star magnetosphere near the periastron passages, which would therefore cause regular variations in the spin-down trend. A few years later, \citet{2005AdSpR..35.1185D} searched for the orbital period using data from the Rossi X-Ray Timing Explorer (RXTE) and the Burst and Transient Source Experiment (BATSE), but were unable to confirm the previous findings of 304d. They instead reported a tentative detection of a 420-460d orbital period. Following this, an analysis of infrared radial velocities of the optical companion conducted by \citet{2006ApJ...641..479H} ruled out the candidate period of 304d and instead reported strong evidence for an orbital period of $\sim1161$d alongside an eccentricity of $e=0.101$. In addition to this, \citet{2006ApJ...641..479H} proposed an upper limit of $M=1.22M_\odot$ of the companion, which would correspond to an inclination of the system of up to 90° . They also ruled out a near face-on inclination of the system, mentioning that an eclipse of the neutron star is theoretically possible. \citet{2007AAS...210.2002C} searched for orbital periodicity using a relatively small time interval of \textit{Swift} Burst Alert Telescope (\textit{Swift}/BAT) data spanning approximately 200 days and RXTE All-Sky Monitor (RXTE/ASM) data covering about 4000 days. They reported no significant detection of either the 304d, 420-460d, or 1161d period, however. Analysis of optical I-band data from the Optical Gravitational Lensing Experiment (OGLE) conducted by \citet{2016pas..conf..133M} was similarly unable to verify the detection of the 1161d orbital period. Instead, \citeauthor{2016pas..conf..133M} reported a clear detection of a $295\pm70$d period, which therefore would support the findings of \citet{1986ApJ...300..551C} and \citet{1999ApJ...526L.105P}. In the following year, \citet{2017A&A...601A.105I} conducted a multi-wavelength study of the system using optical data from the Klein Karoo Observatory as well as observations from the International Gamma-Ray Astrophysics Laboratory (INTEGRAL) using data from the Optical Monitoring Camera (OMC) and the INTEGRAL Soft Gamma-Ray Imager (ISGRI) as well as RXTE observations, where no detection of any of the previously mentioned periods was made. Instead, \citeauthor{2017A&A...601A.105I} argued that the flaring behaviour of the source, which would increase in activity near the periastron passages, supports the 1161d period reported by \citet{2006ApJ...641..479H}. To our knowledge, the most recent work supporting the 1161d period reported by \citeauthor{2006ApJ...641..479H} and \citeauthor{2017A&A...601A.105I} was reported by \citet{2017MNRAS.469.2509S}. In this work, the authors found long-term variations of the torque-luminosity correlation of the source at timescales similar to the 1161d orbital period. Additionally, their timing analysis indicated a strong noise level at the orbital timescales that can even dominate the frequency residuals for the suggested orbital semi-major axis. The period of the system is today generally accepted as $1160.8\pm12.4$d following the radial velocity measurements of \citet{2006ApJ...641..479H} because they provided more compelling evidence for orbital motions than other methods. A conclusive detection of the orbital period in X-rays still remains to be made, however.

We use multiple long-baseline X-ray observations of GX 1+4 from different missions in order to independently determine the orbital period of the system. To do this, we make use of Lomb-Scargle and epoch-folding periodograms and search directly for periodicity in light-curve and spin-frequency data of the source. Furthermore, we support our findings using folded light curves and hardness ratios as well as images of the source during the predicted eclipse.

\section{Observations and data selection}
We used all available public INTEGRAL/ISGRI observations from 52698 MJD to 59694 MJD with an off-axis angle of $\leq 10°$ in the energy range of 28 to 40 keV, and all INTEGRAL Joint European X-Ray Monitor (INTEGRAL/JEM-X) observations from 53052 MJD to 60009 MJD with an off-axis angle of $\leq 3.5°$ in the energy ranges from 4-10 keV and 10-20 keV, respectively.
The data were reduced using the Multi-Messenger Online Data Analysis (MMODA) service\footnote{\url{https://www.astro.unige.ch/mmoda/}} on OSA, version 11.2. The time bin for the ISGRI and JEM-X data reduction was set to 1000s. Since the two JEM-X telescopes do not always operate simultaneously, data from the two detector units were combined. Since the JEM-X data were not suitable for direct analysis due to numerous gaps in the data and a low number of data points that resulted in a poor coverage of the orbit, we instead only used it to support our findings on multiple occasions throughout the analysis.

Additionally, we used \textit{Swift}/BAT daily light curves in the 15-50 keV energy range between 53419 MJD and 59142 MJD that were taken directly from the web-archive\footnote{\url{https://swift.gsfc.nasa.gov/results/transients/index.html}} (see also \citealt{2013ApJS..209...14K}). Data points after 59142 MJD were excluded due to strong noise that was likely caused by instrument degradation.

Furthermore, we used light curves from the Monitor of All-sky X-ray Image (MAXI) \citep{2009PASJ...61..999M} in the energy ranges of 5-10 keV, 10-20 keV, and 5-20 keV from 55054 MJD to 59885 MJD. We decided to deviate from the standard MAXI light curves, which usually start at 4 keV, due to an apparent excess in background radiation below 5 keV visible in the MAXI spectra, and we excluded data below this value. The data reduction was performed using the MAXI on-demand processing\footnote{\url{http://maxi.riken.jp/mxondem/index.html}} online service.

To supplement the light-curve data, we used all available pulse frequency measurements of the neutron star in the spin-down regime between 46885 MJD and 59864 MJD. These measurements consist of data that were taken by multiple missions and can be found in Appendix B in \citet{2012A&A...537A..66G} in addition to spin-frequency measurements from \citet{2017MNRAS.469.2509S}, \citet{2004ApJ...602..320C} and the Fermi Gamma-ray Burst Monitor (Fermi/GBM) archive\footnote{\url{https://gammaray.nsstc.nasa.gov/gbm/science/pulsars/lightcurves/gx1p4.html}}.

\section{Data analysis and results}
\subsection{Orbital period}
\subsubsection{Light curves}
For each of the available data sets, we calculated the Lomb-Scargle periodogram using the LombScargle routine implemented through the astropy\footnote{\url{https://docs.astropy.org/en/stable/timeseries/lombscargle.html}} \citep{2022ApJ...935..167A} package in Python 3.9, taking measurement uncertainties into account (see \citet{2012cidu.conf...47V, 2015ApJ...812...18V} for the methods used in the astropy package). The period corresponding to the highest power in the periodogram are considered to be an indicator of the orbital periodicity $P_\mathrm{orb}$.

We then calculated the false-alarm level corresponding to a false-alarm probability of 0.1\%. The false-alarm level indicates the probability that a peak of the corresponding power was caused by random fluctuations in the data set, under the assumption that the data follow a purely Gaussian distribution (i.e. do not contain a signal) (see \citealt{2018ApJS..236...16V}). While this is a common way to ensure the significance of a peak, other peaks would also be significant, such as harmonics at fractions of the periodicity, or periodicities caused by other factors that might be artificially enhanced by red noise at lower frequencies.
Furthermore, the false-alarm probability prevented us from drawing conclusions regarding the uncertainty on the associated period. One way to do this would be to employ a bootstrapping approach. However, the Lomb-Scargle periodogram was not sensitive enough to the changes introduced by a bootstrapping approach, likely due to the strong inherent variability of the data, which interplays with any periodic signal. We therefore needed to use a different approach to determine the uncertainties on the found periods. 
This was done by first folding the light curve using the found period and then calculating the orbital phase of each point in the data set. The reference epoch $T_0$ for the folding was chosen to be 54557.5 MJD, which was extrapolated based on the time of the predicted eclipse ($T_\mathrm{conj}=52235.3$ MJD) provided by \citet{2006ApJ...641..479H} to be closer to the beginning of our data sets. Following this, we interpolated over the folded profile to obtain the expected amplitude of the signal corresponding to each data point and subtracted this expected signal from the original data set. We then confirmed that the peak in the periodogram created from the subtracted data set corresponding to the previously found period had vanished or was below the detection threshold.

Next, we generated signals with a number of trial periods $P_{\mathrm{trial},i}$ in a range of $\pm 200$d around the previously found period. For each trial period $P_{\mathrm{trial},i}$, we calculated the phases $\Phi_j$ of each data point in our data set as
\begin{equation}
    \Phi_j = \frac{t_j - T_0}{P_{\mathrm{trial},i}}.
\end{equation}
We then calculated the expected rate $r_j$ for each point in phase $\Phi_j$ using the interpolation we obtained earlier, and we replicated the shape of the original signal in this way, but with a varying periodicity.
Based on our assumption of the intrinsic variability of the data shifting the peak of the periodogram, this new signal was then re-added to the subtracted data set with a random phase shift $\Delta \phi$. The Lomb-Scargle periodogram was then calculated, and the period corresponding to the peak with the highest power was saved. This process was repeated 1000 times for each trial period in order to create a distribution of found periods for each inserted signal with a fixed trial period. We then calculated the differences between the periodicity of the inserted signal and the detected periods. The resulting distributions do not resemble a normal distribution in all cases, which means that the standard deviation cannot be used as a 68\% confidence interval. Instead, we chose to iteratively calculate the intervals of each distribution centred around the mean value, which contains 34\% of the periods to the left and right of the mean value, respectively, which then represent the asymmetric uncertainties on the obtained periods.

Next, we searched for a periodicity in the data using epoch-folding. This approach relies on folding the data over a range of trial periods and maximising the resulting $\chi^2$ statistic. Profiles were obtained by applying good time interval (GTI) exposure correction prior to folding. We applied this technique to the available light curves and summarise the results in Table \ref{tab:periods_lightcurves}, alongside the results of the Lomb-Scargle periodograms. In this case, the uncertainties were obtained by bootstrapping.

\begin{figure}[!h]
    \centering
    \includegraphics[width=0.9\columnwidth]{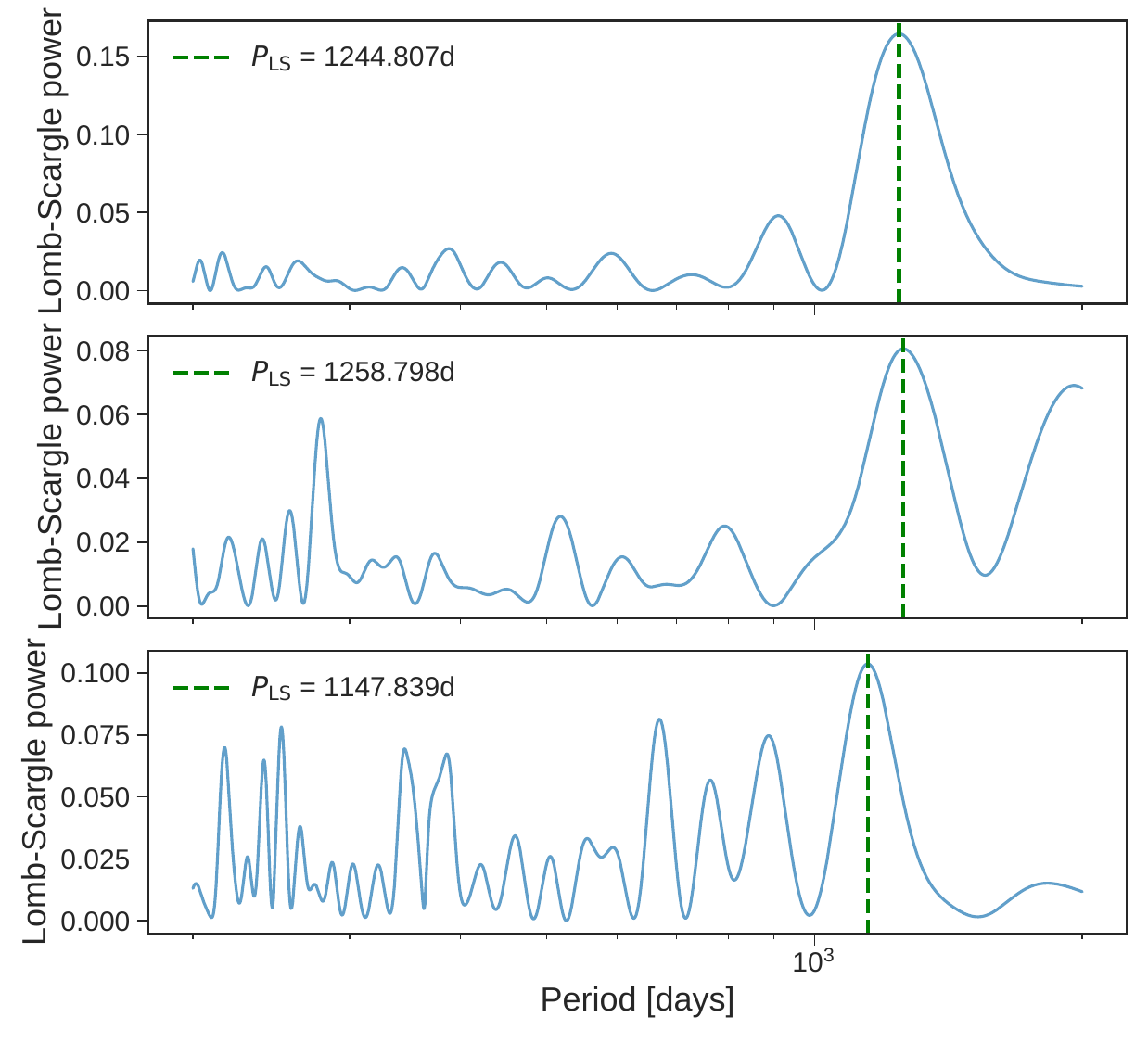}
      \caption{Lomb-Scargle periodograms of the light curves. \textit{Swift} (top), MAXI (middle), and ISGRI (bottom).}
         \label{fig:lomb_scargle_lightcurve}
\end{figure}

\begin{figure}[!h]
    \centering
    \includegraphics[width=0.9\columnwidth]{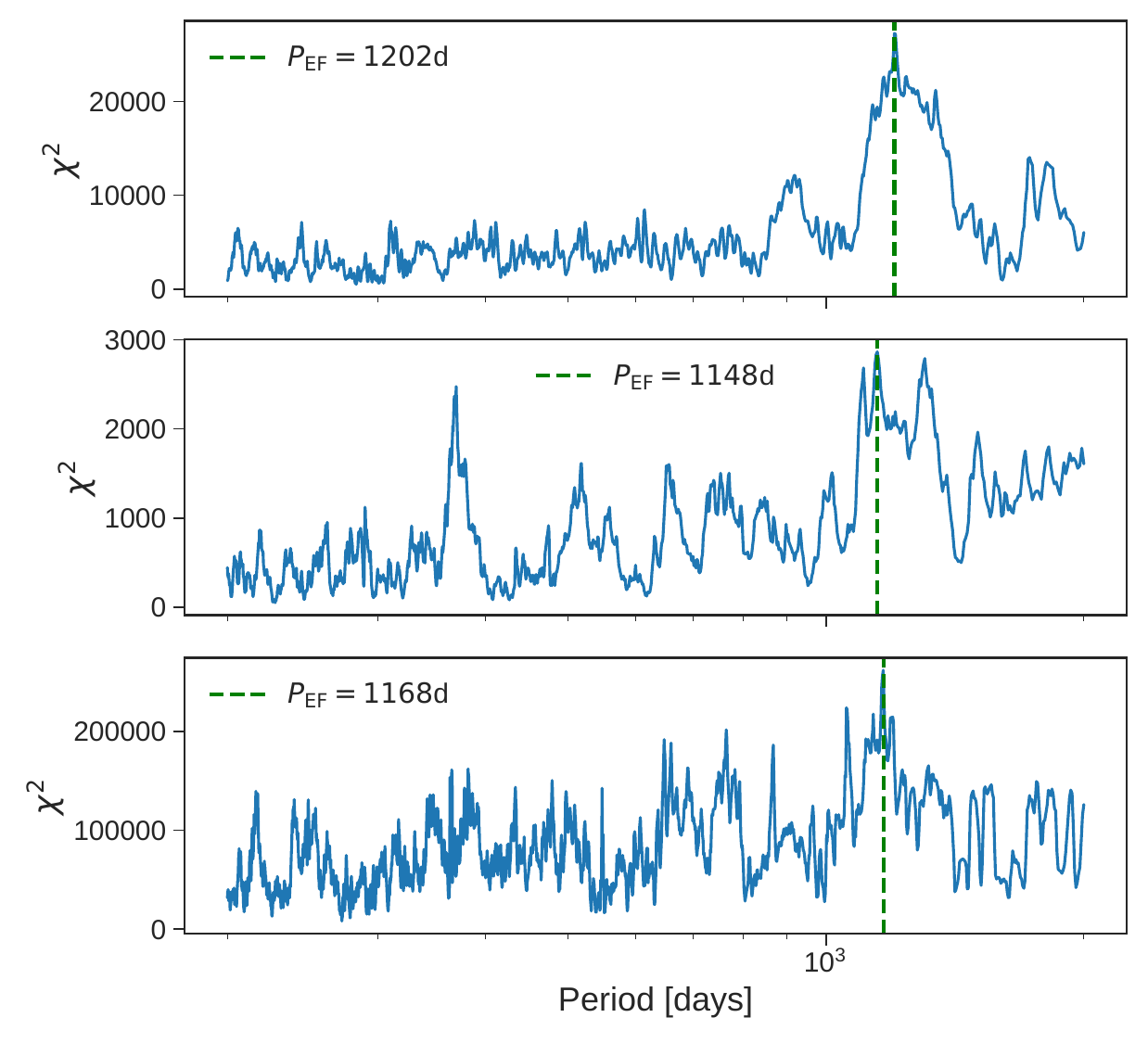}
      \caption{Epoch-folding periodograms of the light curves. \textit{Swift} (top), MAXI (middle), and ISGRI (bottom).}
         \label{fig:epoch_folding_lightcurve}
\end{figure}

\begin{figure*}[!h]
    \sidecaption
    \includegraphics[width=11.3cm]{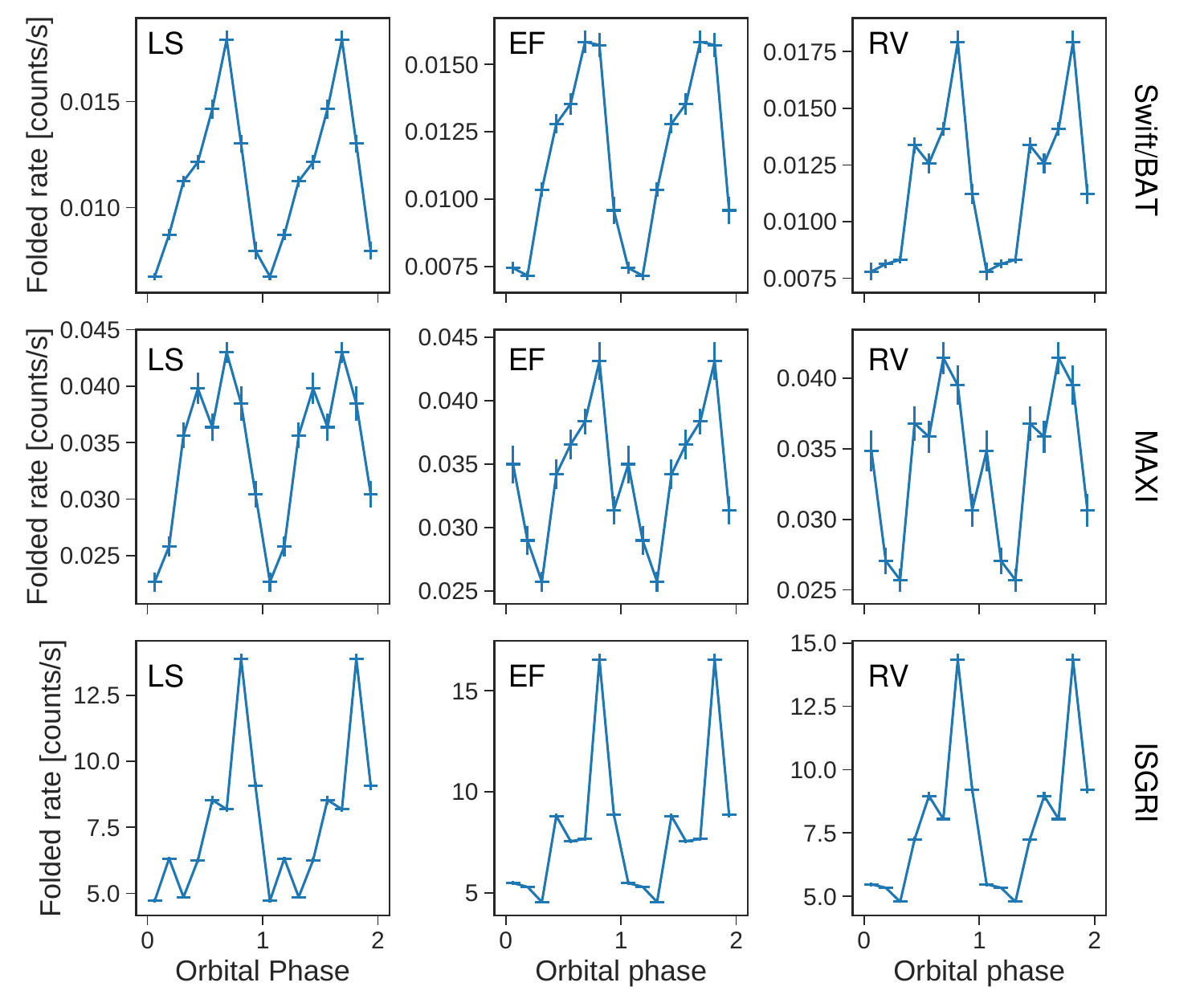}
      \caption{Folded light curves. We show folded \textit{Swift} (first row), MAXI (second row), and ISGRI (third row) light curves with the folding periods being the corresponding Lomb-Scargle (LS) period $P_\mathrm{LS}$ (left column), the epoch-folding  (EF) period $P_\mathrm{EF}$ (centre column), and the radial velocity (RV) period $P_\mathrm{RV}$ (right column) from \citet{2006ApJ...641..479H}. The phase bin around phase 0 corresponds to the phase bin of the predicted eclipse.}
         \label{fig:folded_lightcurves}
\end{figure*}

The Lomb-Scargle and epoch-folding periodograms are shown in Fig.\ref{fig:lomb_scargle_lightcurve} and Fig.\ref{fig:epoch_folding_lightcurve}, respectively. Light curves folded with the Lomb-Scargle period, the epoch-folding period, and the radial velocity period from \citet{2006ApJ...641..479H} are shown in Fig.\ref{fig:folded_lightcurves}. As the reference epoch $T_0$ for the folding, we again used 54557.5 MJD. Since this time was extrapolated based on the radial velocity ephemerides by \citet{2006ApJ...641..479H} and corresponds to the time of the predicted eclipse, it is important to note that the phase bin around phase zero of the folded profiles therefore likewise represents the phase bin of the predicted eclipse.
\subsubsection{Hardness ratio}
\begin{figure*}[!h]
    \sidecaption
    \includegraphics[width=11.3cm]{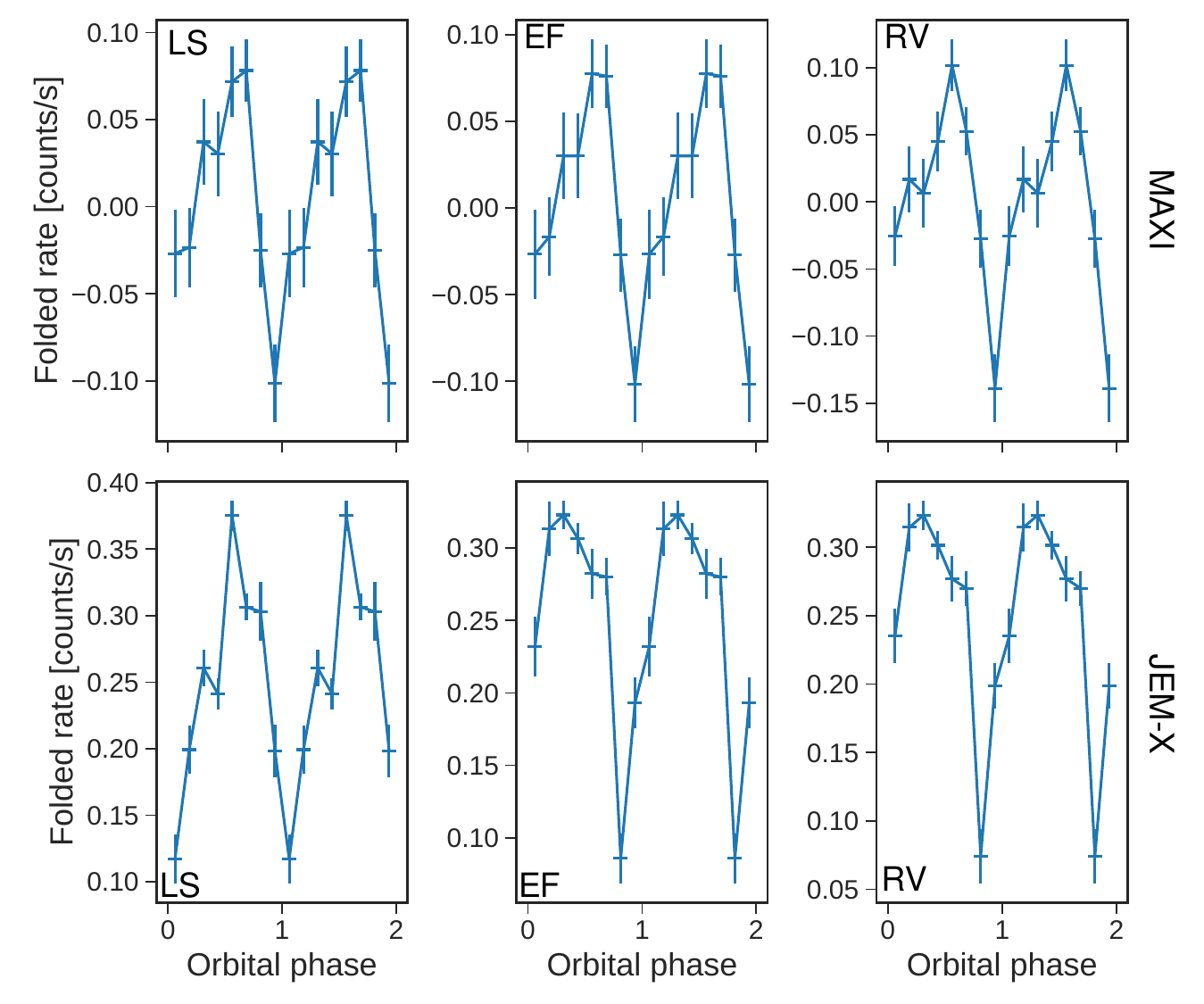}
      \caption{Folded hardness ratios. We show MAXI (first row) and JEM-X (second row) hardness ratios folded with the Lomb-Scargle (LS) period $P_\mathrm{LS}$ (left column), the epoch-folding (EF) period $P_\mathrm{EF}$ (centre column), and the radial velocity (RV) period $P_\mathrm{RV}$ (right column) from \citet{2006ApJ...641..479H}. The phase bin around phase zero corresponds to the phase bin of the predicted eclipse.}
         \label{fig:folded_hardness}
\end{figure*}

In addition to analysing the light curves, we also calculated the hardness ratio, which we defined as 
\begin{equation}
    \mathrm{HR} = \frac{F_\mathrm{soft}-F_\mathrm{hard}}{F_\mathrm{soft}+F_\mathrm{hard}}
\end{equation}
for the 5-10 keV and 10-20 keV bands of MAXI and JEM-X data. Then, we searched for periodicities in the hardness ratios using the methods outlined in the previous section. The JEM-X data were only used to fold the obtained hardness ratio with the candidate periods in order to check for consistency with the results obtained using MAXI data. The MAXI period is listed in Table \ref{tab:periods_lightcurves}. Following this, we folded the hardness ratios with the obtained periods as well as the 1160.8d radial velocity period $P_\mathrm{RV}$ from \citet{2006ApJ...641..479H}, and we compare the results in Fig.\ref{fig:folded_hardness}. For consistency with the folded light curves in the previous section, the reference epoch $T_0$ for the folding was also chosen to be 54557.5 MJD.

The results presented in Fig. \ref{fig:folded_hardness} suggest that the hardness ratio changes periodically and shows a minimum at the predicted eclipse, which could be due to a modulation in the column density $N_\mathrm{H}$ along the line of sight. We attempted to perform orbital-phase resolved spectroscopy of the system using INTEGRAL as well as MAXI spectra in order to investigate whether the spectral parameters change with the orbital phase, as the hardness ratio suggests. However, due to the degeneracy between the column density and photon index, we were unable to draw definitive conclusions. We then performed a spectral analysis on archival Swift/XRT data in PC mode around the orbital phases 0 (OBSIDs 00050750001, 00050751001, 00050751002, 00050751003, 00050751004, 00050752001, 00050752002, and 00050752003 for a combined 57 ks), 0.15 (OBSID 00081653001 with 1.7 ks) and 0.35 (OBSID 00089914001 with 1.9 ks) using a simple absorbed power-law model ({\tt TBabs*powerlaw} in \textit{XSPEC}), which was sufficient to fit the spectra well\footnote{The XRT analysis was done using the Swift-XRT builder: \url{https://www.swift.ac.uk/user_objects/}, which processed the data using HEASOFT v6.32}. The results of this analysis are presented in Table \ref{tab:spectral_results}. We find that the column density increases by a factor of 10 near the expected eclipse.
\begin{table}[!h]
    \centering
        \caption{Orbital periods from light curves and hardness ratios.}
    \begin{tabular}{ccc}
        \hline
        \noalign{\smallskip}
        Data set & $P_\mathrm{LS}$ (d)& $P_\mathrm{EF}$ (d)\\
        \noalign{\smallskip}
        \hline
        \noalign{\smallskip}
        \smallskip
        $\mathrm{Swift}_\mathrm{LC}$ &  1246$^{+35}_{-48}$  &  1202(32)\\
        \smallskip
        $\mathrm{MAXI}_\mathrm{LC}$ &  1259$^{+51}_{-60}$  &  1148(36)\\
        \smallskip
        $\mathrm{ISGRI}_\mathrm{LC}$ &  1147$^{+7}_{-5}$  &  1168(24)\\
        \smallskip
        $\mathrm{MAXI}_\mathrm{HR}$&  1110$^{+62}_{-89}$  &  1139(34) \\
        \hline
        \smallskip
    \end{tabular}
    \tablefoot{The first column lists the data-set that was searched for a periodicity. The light curves and hardness ratios are indicated by the subscripts LC and HR, respectively. The second and third column show the periods obtained using the Lomb-Scargle ($P_\mathrm{LS}$) and epoch-folding ($P_\mathrm{EF}$) method. The quoted uncertainties are given at the 68\% confidence level.}
    \label{tab:periods_lightcurves}
\end{table}

\begin{table*}[!h]
    \centering
        \caption{Results of the spectral analysis using Swift/XRT data.}
    \begin{tabular}{cccc}
        \hline
        \noalign{\smallskip}
        Parameter & Orbital phase 0 & Orbital phase 0.15 & Orbital phase 0.35\\
        \noalign{\smallskip}
        \hline
        \noalign{\smallskip}
        \smallskip
        Start time [MJD] & 53416.8596 &  57309.3916  &  60507.7666\\
        \smallskip
        End Time [MJD]& 53425.7042 &  57309.4113  & 60507.7886\\
        \smallskip
        Exposure [ks]& 57 &  1.7  &  1.9\\
        \smallskip
        $N_\mathrm{H}$ [$10^{22}\mathrm{cm}^{-2}$] & 47.8$^{+1.7}_{-1.7}$ &  3.5$^{+1.1}_{-1.0}$  & 3.99$^{+0.51}_{-0.48}$\\
        \smallskip
        Photon Index $\Gamma$ & 1.74$^{+0.12}_{-0.12}$ &  0.6$^{+0.3}_{-0.3}$  &  0.59$^{+0.14}_{-0.13}$\\
        \smallskip
        Powerlaw Norm &  0.029$^{+0.006}_{-0.004}$ &  0.014$^{+0.008}_{-0.005}$  &  0.03$^{+0.007}_{-0.005}$\\
        \smallskip
        0.3-10 keV Flux [$10^{-11}$erg/$\mathrm{cm^2}$/s] & 2.61$^{+0.01}_{-0.17}$ & 25.6$^{+0.3}_{-10.3}$  &  74.8$^{+2.1}_{-20.4}$\\
        \smallskip
        cstat & 727.33 &  153.00  & 484.59\\
        \smallskip
        d.o.f. & 723 & 160  & 519\\
        \hline
        \smallskip
    \end{tabular}
    \tablefoot{The first column lists the relevant parameters of the spectral analysis. The remaining columns list the corresponding results of the analysis of the observations at the three orbital phases. The quoted uncertainties are given at the 68\% confidence level.}
    \label{tab:spectral_results}
\end{table*}

\subsubsection{Spin-frequency and spin-frequency derivative evolution}
\begin{figure}[!h]
    \centering
    \includegraphics[width=0.9\columnwidth]{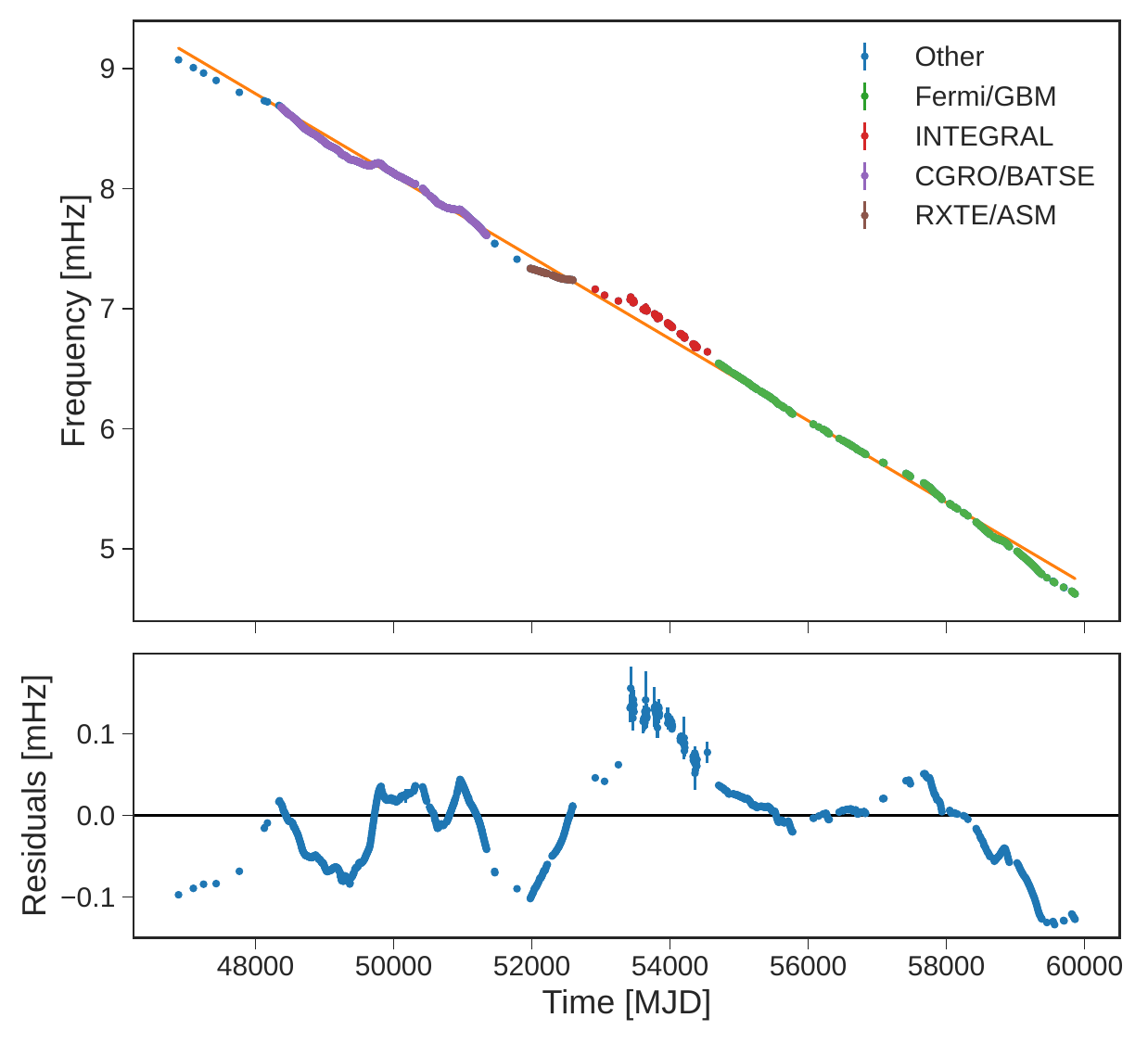}
    \caption{Spin-frequency history of GX 1+4. The data were taken from multiple missions. The solid line represents the linear fit to the data. We searched for periodicity in the residuals in the lower plot.}
    \label{fig:spin_history}
\end{figure}

\begin{figure}[!h]
    \centering
    \includegraphics[width=0.9\columnwidth]{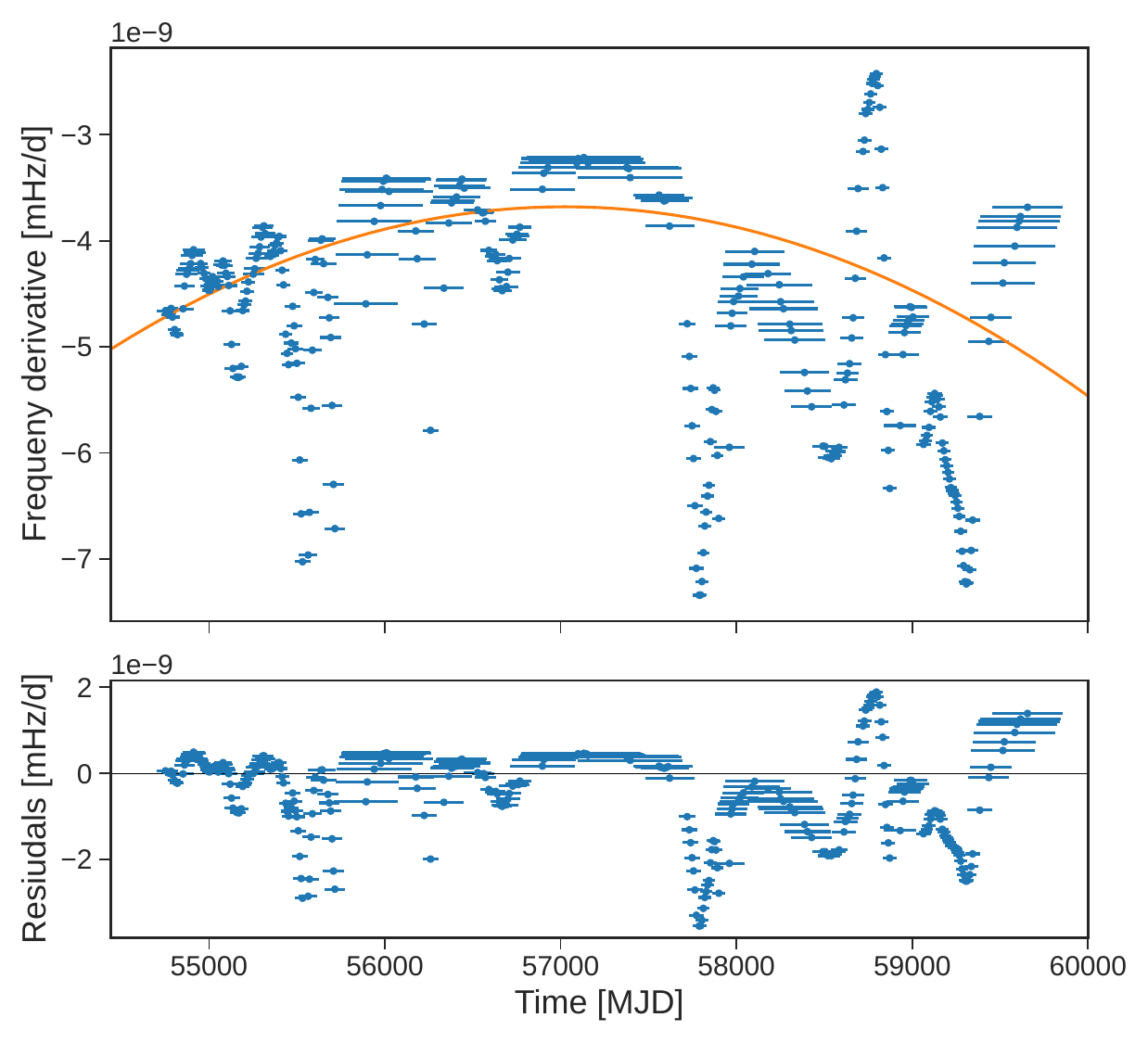}
    \caption{Spin-frequency derivatives of GX 1+4 using Fermi/GBM data. The data points were obtained using a sliding-window approach, and a long-term trend was subtracted (solid orange line; see text).}
    \label{fig:freq_derivatives}
\end{figure}

Lastly, we analysed the spin frequency and spin-frequency derivative evolution of the source to search for a possible periodicity, for example caused by the orbital modulation of the spin frequency or a change in the accretion behaviour near periastron. While this has previously been attempted using BATSE data covering about 3000 days by \citet{1999ApJ...526L.105P}, we instead achieved a baseline of about 12000 days using the combined data from multiple missions, shown in Fig. \ref{fig:spin_history}, which would correspond to a coverage of about ten orbital cycles of the source based on the radial velocity period derived by \citet{2006ApJ...641..479H}.
We first subtracted a linear long-term trend from the spin-frequency evolution of the source and then calculated the Lomb-Scargle periodogram of the subtracted spin-frequency evolution. Similar to the previous sections, we also applied the epoch-folding technique to the data. The results of the two approaches are summarised in Table \ref{tab:spin_history_periods}.
Although the Lomb-Scargle periodogram of the data suggests a periodicity that is consistent with the radial velocity period, the corresponding peak lies below the 0.1\% false-alarm level threshold, and we furthermore could not apply the approach to determine the uncertainties outlined in the previous section since removing the folded signal did not remove the peak in the periodogram. There is an inherent variability in the spin evolution that is associated with variations in accretion torques, which might be the cause for the above. Therefore, we cannot report any uncertainty on the Lomb-Scargle period and only used this data set to support our claims.
In addition to the spin-frequency history, we also searched for a periodicity in the spin-frequency derivatives of the source. To do this, we restricted our analysis to data from Fermi/GBM between 54704 MJD and 59864 MJD because the coverage is better, since measuring a single frequency derivative corresponding to large gaps may smear out the periodic signals. We employed a sliding-window approach where we selected ten consecutive data points to which we fit a linear function with the slope corresponding to the spin frequency derivative. We then moved the window one data point forward in time and repeated the fit until we reached the end of the data set. Since the Lomb-Scargle periodogram of the spin-frequency derivatives showed a strong peak at around 3600 days, we fit a quadratic function to the data, which we considered to be a part of the long-term intrinsic evolution due to accretion (shown in Fig.\ref{fig:freq_derivatives}), and subtracted it from the spin-frequency derivatives to remove the underlying trend and performed the analysis on the subtracted data set. As before, we also analysed the spin-frequency derivatives using the epoch-folding technique and summarise the results in Table \ref{tab:spin_history_periods}.
\begin{table}[!h]
    \centering
        \caption{Orbital periods obtained from the spin frequency and spin-frequency derivative history.}
    \begin{tabular}{ccc}
        \hline
        \noalign{\smallskip}
        Data set & $P_\mathrm{LS}$ (d)& $P_\mathrm{EF}$ (d)\\
        \noalign{\smallskip}
        \hline
        \noalign{\smallskip}
        \smallskip
        $\nu_\mathrm{spin}$ &  1124(-) &  1178(-) \\
        \smallskip
        $\dot{\nu}_\mathrm{spin}$ & 1194$^{+20}_{-76}$  & 1212(37) \\
        \hline
        \smallskip
    \end{tabular}
    \tablefoot{The first column lists the data set that was searched for periodicity. The spin frequency and spin-frequency derivative history are denoted by $\nu_\mathrm{spin}$ and $\dot{\nu}_\mathrm{spin}$, respectively. The second and third column list the results obtained using the Lomb-Scargle ($P_\mathrm{LS}$) and epoch-folding ($P_\mathrm{EF}$) method. The quoted uncertainties are given at the 68\% confidence level.}
    \label{tab:spin_history_periods}
\end{table}
\subsection{INTEGRAL images}
In contrast to the periodicities that are deduced from X-ray variabilities through periodograms, the radial velocity measurements from \citet{2006ApJ...641..479H} provide a total orbital solution, including the amplitude of the semi-major axis and the time of periastron passages, as well as the time $T_0$ of a possible eclipse of the neutron star. Therefore, we used this information to search for an eclipse in the X-ray data.

To do this, we created ISGRI and JEM-X mosaic images for the orbital phase bin around the orbital phase corresponding to the expected position of the eclipse as predicted by \citet{2006ApJ...641..479H}. Since we extrapolated the reference epoch for folding from the radial velocity ephemerides, eclipses should in principle repeat at each passage $n\cdot T_0$, leading to a minimum in the folded profile when folded with the given ephemeris. However, since the uncertainty projection after many orbital cycles leads to an ambiguity in the arrival times of the eclipses and because we folded the profiles at a resolution of eight bins (corresponding to a coverage of 145.125d each), we also considered the possibility of an occurrence of the eclipse in adjacent phase bins. In addition to the phase-resolved imaging of the source, we also performed imaging on an orbit-by-orbit basis. We did not find any evidence for a complete eclipse of the neutron star by the companion, however, because the source was clearly detected in every image.

\section{Discussion and conclusions}
The orbital period of GX 1+4 has been a point of contention for a long time and until radial velocity measurements suggested the 1161d period as the orbital period of the system \citep{2006ApJ...641..479H}, which has generally been accepted because the radial velocity measurements provide compelling evidence. This detection has not been reproduced in X-rays so far.

Based on their assumptions on the mass of the neutron star and the value of the mass function, \citet{2006ApJ...641..479H} suggested a maximum mass of the companion of $1.22 M_\odot$ for a neutron star mass of $M_{NS}=1.35M_\odot$, which requires a corresponding inclination of the system of $\sim$ 90° and provides ephemerides for a predicted total eclipse. This means that even in X-rays, a periodic decrease in brightness of the source can be expected, which should be detectable in the long-term light curves of the source. Furthermore, the orbit of the system is eccentric, and changes in the accretion torque at periastron could therefore have an effect on the spin-frequency history.

\begin{figure*}[!h]
    \sidecaption
    \includegraphics[width=11.3cm]{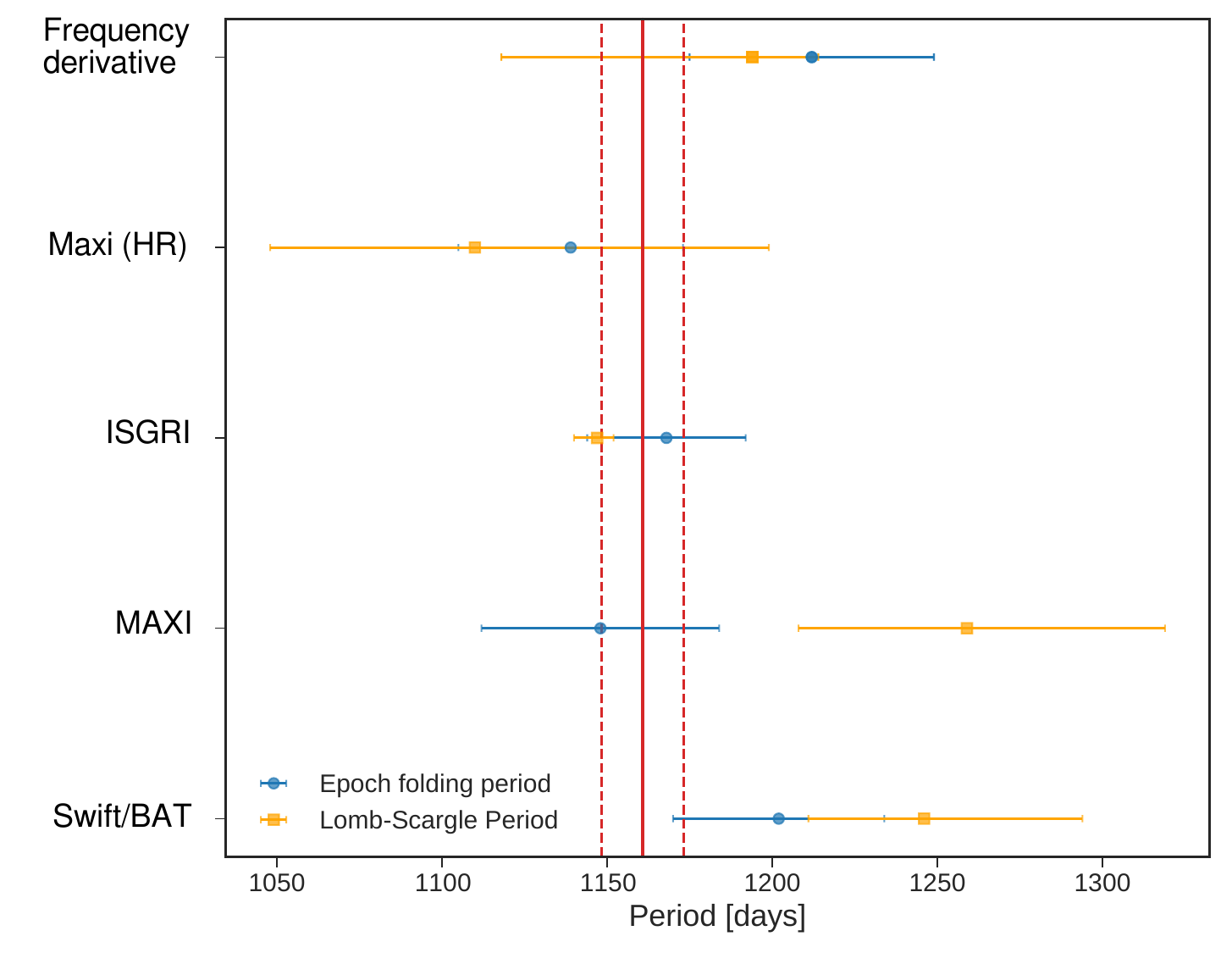}
    \caption{Visual summary of the analysis results. The blue circles represent the results obtained with the epoch-folding approach, and the results obtained with the Lomb-Scargle technique are shown as orange squares. The data sets are Swift, MAXI, ISGRI, MAXI hardness ratio, and spin-frequency derivative evolution, as indicated by the labels in the figure. The solid red line indicates the 1160.8d radial velocity period obtained by \citet{2006ApJ...641..479H}, and the dashed lines mark the corresponding 12.4d uncertainty on this value.}
    \label{fig:visualisation}
\end{figure*}

We searched for a periodicity of GX 1+4 using different approaches and data sets. Lomb-Scargle and epoch-folding periodograms were calculated from the light curves, hardness ratios, and spin-frequency derivative history of the source. An overview of the results is presented in Fig.\ref{fig:visualisation}, and detailed results are presented in Table \ref{tab:periods_lightcurves} and Table \ref{tab:spin_history_periods}. The results obtained using these two independent methods agree with the radial velocity period proposed by \citet{2006ApJ...641..479H} within the 68\% confidence interval.
This implies that the light curves of the source show behaviours that occur periodically on the same timescale as the proposed orbital period, which is consistent with the assumption of an increase in accretion rate, and consequently, the luminosity, near periastron and a decrease in luminosity near the predicted eclipse. 

Furthermore, we folded the light curves and hardness ratios with the radial velocity period and obtained profiles with features that are consistent with the expectations from the radial velocity ephemerides. The light curves folded with the radial velocity period in the rightmost column of Fig.\ref{fig:folded_lightcurves} show an excellent agreement between Swift and ISGRI data. The minima of the profiles coincide within the statistical uncertainties with the predicted eclipse, but are not consistent with zero.  
This suggests the absence of a complete eclipse and may instead be caused by an increase in column density in the line of sight, possibly due to the outer layers of the M-type giant companion entering the line of sight between the neutron star and the observer, leading to an increase in absorption. This is further supported by the folded hardness ratios, which are at a minimum at the expected eclipse, suggesting either an increase in hard X-rays or a decrease in soft X-ray emission via increased absorption due to the companion. The results of the spectral analysis using Swift/XRT data, summarised in Table \ref{tab:spectral_results}, support this assumption of an increased column density $N_\mathrm{H}$ as they show an increase by a factor of 10 near the expected eclipse. However, the results of the spectral analysis do not show a hardening of the spectral shape at the predicted eclipse (orbital phase $\Phi =0$). On the contrary, we observe a slight softening. We also do not observe the typical appearance of a bright iron line complex (see e.g. \citealt{2011ASInC...3Q.138N}). Additionally, we found pulsations in the light curves of the source in some of the observations taken near the eclipse, suggesting that the neutron star is observed directly.
The images obtained with INTEGRAL likewise support the absence of a complete eclipse as the source is consistently detected with a high detection significance ($\gtrapprox 20\sigma$) throughout all the phasevintervals that we analysed.
The folded light curves obtained using MAXI data (rightmost column of Fig.\ref{fig:folded_lightcurves}) show a secondary peak at $\Phi \sim 0$ that is not seen with Swift/BAT and ISGRI. This might be caused by the different energy ranges of MAXI with respect to Swift and INTEGRAL, suggesting an energy dependence of the MAXI profiles. Since we do not see this secondary peak in the folded JEM-X light curve, another reason could be contamination of MAXI data from two neighbouring LMXBs GX 3+1 and SLX 1735-269 due to the large point spread function (PSF) of 1.5° of the instrument.

Since the orbit of the neutron star has a non-zero eccentricity, the accretion torque is likely to vary over the course of the orbit. These changes would therefore be reflected in the spin frequency and spin-frequency derivative history. We find periodicities in the spin frequency and spin-frequency derivative history of the source that would agree with the proposed radial velocity period and could therefore be indicative of these changes in accretion torque. 

\begin{figure*}
    \sidecaption
    \includegraphics[width=11.3cm]{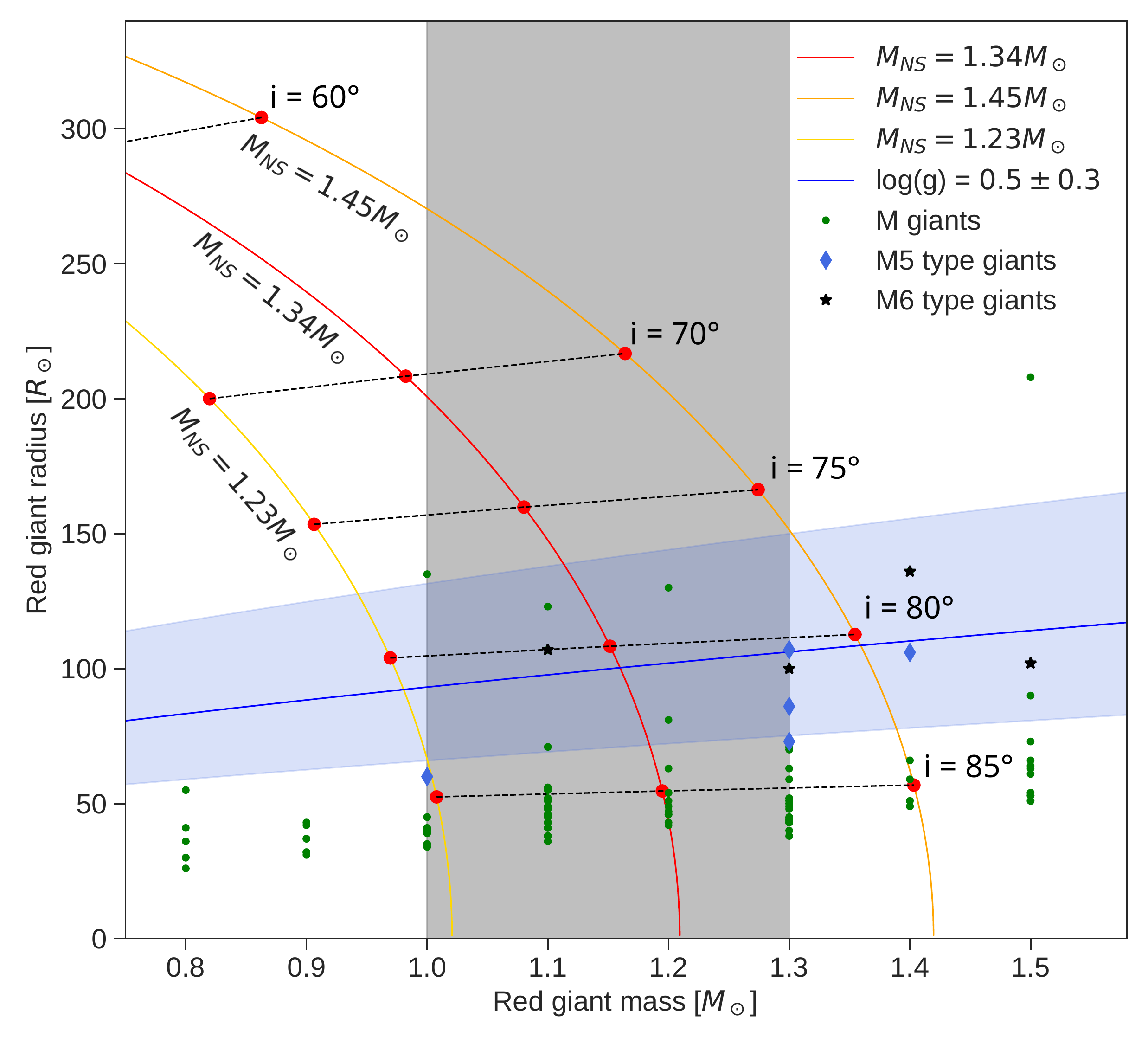}
    \caption{Companion radius as function of mass. The solid lines indicate the limits placed on the radius of the companion at different inclinations for a given neutron star mass. The mass-radius relation of M giants from the results by \citet{1998NewA....3..137D} is indicated by the green dots. M5- and M6-type giants from this sample are represented by blue diamonds and black stars, respectively. The solid blue line represents the radius calculated from the gravitational acceleration $\log{(g)}$ as determined by \citet{2006ApJ...641..479H} for a given red giant mass, and the area shaded in light blue indicates the corresponding uncertainty. The area shaded in grey indicates the mass limits of the red giant as determined by \citet{2006ApJ...641..479H} through a spectroscopic analysis of the oxygen isotopic abundances of $^{17}$O and $^{16}$O.}
    \label{fig:mass_vs_radius}
\end{figure*}

The lack of the eclipse emerging from our analysis implies that the binary system is not observed edge-on ($i \neq 90^\circ$). We show that this result, together with other information obtained by \citet{2006ApJ...641..479H}, allows us to place constraints on the inclination $i$ of the orbital plane and the mass of the neutron star. We first calculated the red giant radius as a function of the red giant mass using the surface gravity provided by \citet{2006ApJ...641..479H} ($\log{(g)} = 0.5\pm0.3$). The resulting region of allowed masses and radii is represented by the blue line and light blue region in Fig. \ref{fig:mass_vs_radius}. Furthermore, based on the abundances of $^{17}$O and $^{16}$O, \citet{2006ApJ...641..479H} determined a range of allowed masses for the red giant ranging from $\sim1M_\odot$ to $\sim1.3M_\odot$. This is represented by the grey region in Fig. \ref{fig:mass_vs_radius}. The region in which the grey and blue regions overlap represents the possible solutions for the mass and radius of the red giant.

Following this, we calculated the resulting masses of the companion and the semi-major axes of the system for a number of different neutron star masses and inclinations between 40° and 90° using the results for orbital period and the mass function of the system provided by the radial velocity measurements of \citet{2006ApJ...641..479H}.
By calculating the separation of the companion and the neutron star at the time of the predicted eclipse (orbital phase $\Phi=0$), we were able to obtain a maximum radius that the red giant can have in order to not eclipse the neutron star for different neutron star masses and inclinations. This is shown in Fig.\ref{fig:mass_vs_radius}. Since we do not find any evidence for a complete eclipse of the system, the radius of the red giant must be smaller than the value indicated for the red giant in Fig. \ref{fig:mass_vs_radius}, corresponding to the given inclination for any specified $M_\mathrm{NS}$ (i.e. below the red, orange, or yellow lines). 
The intersection of the grey area with the light blue area in Fig. \ref{fig:mass_vs_radius} furthermore allows us to place constraints on the neutron star mass and the inclination of the system. 
Based on this intersection, we find a range of neutron star masses of $\sim1.23M_\odot$ to $\sim1.45M_\odot$ and inclinations of the system in the range of $\sim76$° to $\sim84$°.

Considering that the spectral type of the companion in the system was determined to be either M6 III \citep{1977ApJ...211..866D}, or M5 III \citep{1997ApJ...489..254C}, we can compare our results to the findings of \citet{1998NewA....3..137D} for masses and radii of nearby M giants. This is represented by the green dots in Fig.\ref{fig:mass_vs_radius}, where M5 and M6 giants are distinguished from this sample by blue diamonds and black stars, respectively. Since all allowed solutions for the red giant in the system have to be contained within the overlap region between the grey and blue regions, we can immediately see that some of the red giants from the sample by \citet{1998NewA....3..137D} fall within that area.

However, the calculations leading to this result assumed that in the low state at the predicted eclipse, the hard X-ray emission originates directly from the neutron star rather than being scattered by an intervening medium such as the wind of the red giant, in which case we would expect to observe spectral variability and characteristic emission lines (see e.g. \citealt{2011ASInC...3Q.138N}, \citealt{2003ApJ...582..959W}).
The sparse Swift/XRT data indicate an increase in the column density around the expected eclipse (orbital phase $\Phi=0$), which could be a possible cause for the observed hardness ratio modulation, but it remains unclear whether the spectral continuum is also periodically variable. Future observations of the entire orbit, covering both soft (<1 keV) and hard X-ray energies, are crucial to fully understand the implications of the spectral analysis and the consequences of the absence of an eclipse in the system.

\begin{acknowledgements}
Based on observations with INTEGRAL, an ESA project with instruments and science data centre
funded by ESA member states (especially the PI countries: Denmark, France, Germany, Italy,
Switzerland, Spain), and with the participation of the Russian Federation and the USA. 
We acknowledge the use of public data from the \textit{Swift} data archive.
This research has made use of MAXI data provided by RIKEN, JAXA and the MAXI team.
This paper is supported by European Union's 2020 research and innovation programme under grant agreement No.871158, project AHEAD2020.
\end{acknowledgements}

%
%
\bibliographystyle{aa} 
\bibliography{bibliography} 

\begin{thebibliography}{32}
\expandafter\ifx\csname natexlab\endcsname\relax\def\natexlab#1{#1}\fi

\bibitem[{{Astropy Collaboration} {et~al.}(2022){Astropy Collaboration}, {Price-Whelan}, {Lim}, {Earl}, {Starkman}, {Bradley}, {Shupe}, {Patil}, {Corrales}, {Brasseur}, {N{\"o}the}, {Donath}, {Tollerud}, {Morris}, {Ginsburg}, {Vaher}, {Weaver}, {Tocknell}, {Jamieson}, {van Kerkwijk}, {Robitaille}, {Merry}, {Bachetti}, {G{\"u}nther}, {Aldcroft}, {Alvarado-Montes}, {Archibald}, {B{\'o}di}, {Bapat}, {Barentsen}, {Baz{\'a}n}, {Biswas}, {Boquien}, {Burke}, {Cara}, {Cara}, {Conroy}, {Conseil}, {Craig}, {Cross}, {Cruz}, {D'Eugenio}, {Dencheva}, {Devillepoix}, {Dietrich}, {Eigenbrot}, {Erben}, {Ferreira}, {Foreman-Mackey}, {Fox}, {Freij}, {Garg}, {Geda}, {Glattly}, {Gondhalekar}, {Gordon}, {Grant}, {Greenfield}, {Groener}, {Guest}, {Gurovich}, {Handberg}, {Hart}, {Hatfield-Dodds}, {Homeier}, {Hosseinzadeh}, {Jenness}, {Jones}, {Joseph}, {Kalmbach}, {Karamehmetoglu}, {Ka{\l}uszy{\'n}ski}, {Kelley}, {Kern}, {Kerzendorf}, {Koch}, {Kulumani}, {Lee}, {Ly}, {Ma}, {MacBride}, {Maljaars}, {Muna}, {Murphy}, {Norman},
  {O'Steen}, {Oman}, {Pacifici}, {Pascual}, {Pascual-Granado}, {Patil}, {Perren}, {Pickering}, {Rastogi}, {Roulston}, {Ryan}, {Rykoff}, {Sabater}, {Sakurikar}, {Salgado}, {Sanghi}, {Saunders}, {Savchenko}, {Schwardt}, {Seifert-Eckert}, {Shih}, {Jain}, {Shukla}, {Sick}, {Simpson}, {Singanamalla}, {Singer}, {Singhal}, {Sinha}, {Sip{\H{o}}cz}, {Spitler}, {Stansby}, {Streicher}, {{\v{S}}umak}, {Swinbank}, {Taranu}, {Tewary}, {Tremblay}, {de Val-Borro}, {Van Kooten}, {Vasovi{\'c}}, {Verma}, {de Miranda Cardoso}, {Williams}, {Wilson}, {Winkel}, {Wood-Vasey}, {Xue}, {Yoachim}, {Zhang}, {Zonca}, \& {Astropy Project Contributors}}]{2022ApJ...935..167A}
{Astropy Collaboration}, {Price-Whelan}, A.~M., {Lim}, P.~L., {et~al.} 2022, \apj, 935, 167

\bibitem[{{Bozzo} {et~al.}(2022){Bozzo}, {Romano}, {Ferrigno}, \& {Oskinova}}]{2022MNRAS.513...42B}
{Bozzo}, E., {Romano}, P., {Ferrigno}, C., \& {Oskinova}, L. 2022, \mnras, 513, 42

\bibitem[{{Chakrabarty} \& {Roche}(1997)}]{1997ApJ...489..254C}
{Chakrabarty}, D. \& {Roche}, P. 1997, \apj, 489, 254

\bibitem[{{Chakrabarty} {et~al.}(1998){Chakrabarty}, {van Kerkwijk}, \& {Larkin}}]{1998ApJ...497L..39C}
{Chakrabarty}, D., {van Kerkwijk}, M.~H., \& {Larkin}, J.~E. 1998, \apjl, 497, L39

\bibitem[{{Corbet} {et~al.}(2007){Corbet}, {Markwardt}, {Mukai}, {Sokoloski}, \& {Tueller}}]{2007AAS...210.2002C}
{Corbet}, R. H.~D., {Markwardt}, C., {Mukai}, K., {Sokoloski}, J., \& {Tueller}, J. 2007, in American Astronomical Society Meeting Abstracts, Vol. 210, American Astronomical Society Meeting Abstracts \#210, 20.02

\bibitem[{{Cui} \& {Smith}(2004)}]{2004ApJ...602..320C}
{Cui}, W. \& {Smith}, B. 2004, \apj, 602, 320

\bibitem[{{Cutler} {et~al.}(1986){Cutler}, {Dennis}, \& {Dolan}}]{1986ApJ...300..551C}
{Cutler}, E.~P., {Dennis}, B.~R., \& {Dolan}, J.~F. 1986, \apj, 300, 551

\bibitem[{{Davidsen} {et~al.}(1977){Davidsen}, {Malina}, \& {Bowyer}}]{1977ApJ...211..866D}
{Davidsen}, A., {Malina}, R., \& {Bowyer}, S. 1977, \apj, 211, 866

\bibitem[{{Dieters} {et~al.}(2005){Dieters}, {O'Neill}, {Farrell}, \& {Sood}}]{2005AdSpR..35.1185D}
{Dieters}, S., {O'Neill}, P., {Farrell}, S., \& {Sood}, R. 2005, Advances in Space Research, 35, 1185

\bibitem[{{Dumm} \& {Schild}(1998)}]{1998NewA....3..137D}
{Dumm}, T. \& {Schild}, H. 1998, \na, 3, 137

\bibitem[{{Glass} \& {Feast}(1973)}]{1973NPhS..245...39G}
{Glass}, I.~S. \& {Feast}, M.~W. 1973, Nature Physical Science, 245, 39

\bibitem[{{Gonz{\'a}lez-Gal{\'a}n} {et~al.}(2012){Gonz{\'a}lez-Gal{\'a}n}, {Kuulkers}, {Kretschmar}, {Larsson}, {Postnov}, {Kochetkova}, \& {Finger}}]{2012A&A...537A..66G}
{Gonz{\'a}lez-Gal{\'a}n}, A., {Kuulkers}, E., {Kretschmar}, P., {et~al.} 2012, \aap, 537, A66

\bibitem[{{Hall} \& {Davelaar}(1983)}]{1983IAUC.3872....1H}
{Hall}, R. \& {Davelaar}, J. 1983, \iaucirc, 3872, 1

\bibitem[{{Hinkle} {et~al.}(2006){Hinkle}, {Fekel}, {Joyce}, {Wood}, {Smith}, \& {Lebzelter}}]{2006ApJ...641..479H}
{Hinkle}, K.~H., {Fekel}, F.~C., {Joyce}, R.~R., {et~al.} 2006, \apj, 641, 479

\bibitem[{{I{\l}kiewicz} {et~al.}(2017){I{\l}kiewicz}, {Miko{\l}ajewska}, \& {Monard}}]{2017A&A...601A.105I}
{I{\l}kiewicz}, K., {Miko{\l}ajewska}, J., \& {Monard}, B. 2017, \aap, 601, A105

\bibitem[{{Jablonski} {et~al.}(1997){Jablonski}, {Pereira}, {Braga}, \& {Gneiding}}]{1997ApJ...482L.171J}
{Jablonski}, F.~J., {Pereira}, M.~G., {Braga}, J., \& {Gneiding}, C.~D. 1997, \apjl, 482, L171

\bibitem[{{Krimm} {et~al.}(2013){Krimm}, {Holland}, {Corbet}, {Pearlman}, {Romano}, {Kennea}, {Bloom}, {Barthelmy}, {Baumgartner}, {Cummings}, {Gehrels}, {Lien}, {Markwardt}, {Palmer}, {Sakamoto}, {Stamatikos}, \& {Ukwatta}}]{2013ApJS..209...14K}
{Krimm}, H.~A., {Holland}, S.~T., {Corbet}, R.~H.~D., {et~al.} 2013, \apjs, 209, 14

\bibitem[{{Lewin} {et~al.}(1971){Lewin}, {Ricker}, \& {McClintock}}]{1971ApJ...169L..17L}
{Lewin}, W. H.~G., {Ricker}, G.~R., \& {McClintock}, J.~E. 1971, \apjl, 169, L17

\bibitem[{{Luna}(2023)}]{2023A&A...676L...2L}
{Luna}, G.~J.~M. 2023, \aap, 676, L2

\bibitem[{{Majczyna} {et~al.}(2016){Majczyna}, {Madej}, {Nale{\.z}yty}, {R{\'o}{\.z}a{\'n}ska}, \& {Udalski}}]{2016pas..conf..133M}
{Majczyna}, A., {Madej}, J., {Nale{\.z}yty}, M., {R{\'o}{\.z}a{\'n}ska}, A., \& {Udalski}, A. 2016, in 37th Meeting of the Polish Astronomical Society, ed. A.~{R{\'o}{\.z}a{\'n}ska} \& M.~{Bejger}, Vol.~3, 133--136

\bibitem[{{Masetti} {et~al.}(2006){Masetti}, {Orlandini}, {Palazzi}, {Amati}, \& {Frontera}}]{2006A&A...453..295M}
{Masetti}, N., {Orlandini}, M., {Palazzi}, E., {Amati}, L., \& {Frontera}, F. 2006, \aap, 453, 295

\bibitem[{{Matsuoka} {et~al.}(2009){Matsuoka}, {Kawasaki}, {Ueno}, {Tomida}, {Kohama}, {Suzuki}, {Adachi}, {Ishikawa}, {Mihara}, {Sugizaki}, {Isobe}, {Nakagawa}, {Tsunemi}, {Miyata}, {Kawai}, {Kataoka}, {Morii}, {Yoshida}, {Negoro}, {Nakajima}, {Ueda}, {Chujo}, {Yamaoka}, {Yamazaki}, {Nakahira}, {You}, {Ishiwata}, {Miyoshi}, {Eguchi}, {Hiroi}, {Katayama}, \& {Ebisawa}}]{2009PASJ...61..999M}
{Matsuoka}, M., {Kawasaki}, K., {Ueno}, S., {et~al.} 2009, \pasj, 61, 999

\bibitem[{{Muerset} {et~al.}(1997){Muerset}, {Wolff}, \& {Jordan}}]{1997A&A...319..201M}
{Muerset}, U., {Wolff}, B., \& {Jordan}, S. 1997, \aap, 319, 201

\bibitem[{{Naik} \& {Paul}(2011)}]{2011ASInC...3Q.138N}
{Naik}, S. \& {Paul}, B. 2011, in Astronomical Society of India Conference Series, Vol.~3, Astronomical Society of India Conference Series, 138

\bibitem[{{Pereira} {et~al.}(1999){Pereira}, {Braga}, \& {Jablonski}}]{1999ApJ...526L.105P}
{Pereira}, M.~G., {Braga}, J., \& {Jablonski}, F. 1999, \apjl, 526, L105

\bibitem[{{Ricketts} {et~al.}(1982){Ricketts}, {Hall}, {Page}, {Whitford}, \& {Pounds}}]{1982MNRAS.201..759R}
{Ricketts}, M.~J., {Hall}, R., {Page}, C.~G., {Whitford}, C.~H., \& {Pounds}, K.~A. 1982, \mnras, 201, 759

\bibitem[{{Serim} {et~al.}(2017){Serim}, {{\c{S}}ahiner}, {{\c{C}}erri-Serim}, {{\.I}nam}, \& {Baykal}}]{2017MNRAS.469.2509S}
{Serim}, M.~M., {{\c{S}}ahiner}, {\c{S}}., {{\c{C}}erri-Serim}, D., {{\.I}nam}, S.~{\c{C}}., \& {Baykal}, A. 2017, \mnras, 469, 2509

\bibitem[{{Tutukov} \& {Iungelson}(1976)}]{1976Afz....12..521T}
{Tutukov}, A.~V. \& {Iungelson}, L.~R. 1976, Astrofizika, 12, 521

\bibitem[{{VanderPlas} {et~al.}(2012){VanderPlas}, {Connolly}, {Ivezic}, \& {Gray}}]{2012cidu.conf...47V}
{VanderPlas}, J., {Connolly}, A.~J., {Ivezic}, Z., \& {Gray}, A. 2012, in Proceedings of Conference on Intelligent Data Understanding (CIDU, 47--54

\bibitem[{{VanderPlas}(2018)}]{2018ApJS..236...16V}
{VanderPlas}, J.~T. 2018, \apjs, 236, 16

\bibitem[{{VanderPlas} \& {Ivezi{\'c}}(2015)}]{2015ApJ...812...18V}
{VanderPlas}, J.~T. \& {Ivezi{\'c}}, {\v{Z}}. 2015, \apj, 812, 18

\bibitem[{{Wojdowski} {et~al.}(2003){Wojdowski}, {Liedahl}, {Sako}, {Kahn}, \& {Paerels}}]{2003ApJ...582..959W}
{Wojdowski}, P.~S., {Liedahl}, D.~A., {Sako}, M., {Kahn}, S.~M., \& {Paerels}, F. 2003, \apj, 582, 959

\end{thebibliography}

\end{document}